\documentclass[aps,twocolumn,prb,showpacs]{revtex4}
\usepackage{graphicx}
\usepackage{epsfig}
\usepackage{amsmath}
\usepackage{graphics}
\begin{document}

\title{Directed motion and useful work from an isotropic nonequilibrium distribution}

\author{M. F. Gelin}
\author{D. S. Kosov}

\affiliation{
Department of Chemistry and Biochemistry, University of Maryland, College Park, MD  20742}

\begin{abstract}
We demonstrate that a gas of classical particles trapped in an external
asymmetric potential undergoes a quasiperiodic motion, if  the temperature of its initial velocity distribution 
$T_{ne}$ differs from the equilibrium temperature, $T_{eq}$. The magnitude of the effect
is determined by the value of $T_{ne}-T_{eq}$, and the direction
of the motion is determined by the sign of this expression. 
The ``loading'' and ``unloading''
of the gas particles  change directions of their motion, thereby creating a possibility of shuttle-like
motion. The system works as a Carnot engine where the heat flow between kinetic and potential parts of the nonequilibrium distribution
produces the useful work.
\end{abstract}


\maketitle
Since the Maxwell demon thought experiment, the 
extraction of useful work and directed motion from unbiased nonequilibrium distributions has been the source of fascination, intrigue and confusion.\cite{landauer} Being a fundamental scientific problem,
it is  also of significant practical interest for various biological and nanotechnological applications. Biological systems driven by the nonequilibrium fluctuations   include translocases, which pull protein across membranes; kinesin, which walks along microtubules; myosin, which moves along actin in muscle; helicases, which unwinds DNA.\cite{kolomeisky} Recently, there have been a flurry of activities to harness nonequilibrium fluctuations to create various types of working nanodevices.\cite{matt,silva} In this Letter, we propose a new type of "motors" which are driven by the heat flow between nonequilibrium velocity and equilibrium coordinate  distributions.

Consider a gas of classical non-interacting particles trapped in external potential $U(x)$. If
the gas is in thermal equilibrium with a heat bath at the temperature
$T_{eq}$, then positions
($x$) and velocities ($v$) of the gas particles have the equilibrium Boltzmann distribution:
\begin{equation}
\rho_{B}(x,v)=\rho_{B}(x)\rho_{B}(v);
\label{Boltz}\end{equation}
\begin{eqnarray}
\rho_{B}(x)&=&\exp\{-U(x)\}/\int dx\exp\{-U(x)\},
\nonumber
\\
\rho_{B}(v)&=&\sqrt{1/(2\pi)}\exp\{-v^{2}/2\}.\label{Bol1}\end{eqnarray}
We use the reduced variables in this Letter: Time, position, velocity,
and energy are measured in units of $\sqrt{mL^{2}/(k_{B}T_{eq}})$,
$L$, $\sqrt{k_{B}T_{eq}/m}$ and $k_{B}T_{eq}$, respectively ($k_{B}$
is the Boltzmann constant, $m$ is the mass of a gas particle and $L$
is a unit of length). 

Since equilibrium fluctuations cannot produce any directed motion,
the average velocity of the gas $\left\langle v(t)\right\rangle_B =\int dv dx \rho_{B} (v,x) v(t)$ 
is identically zero at any time $t$. What happens if we instantaneously 
change the velocity distribution of the particles to some non-equilibrium distribution $\rho_{ne}(v)$?
Apparently, if 
$\rho_{ne}(v)=\delta(v-v')$, then the gas undergoes a periodic motion
within the potential well. Such a distribution (and akin non-singular
distributions) are biased, since they predict nonzero mean initial
velocity $\left\langle v(t=0)\right\rangle =v'$. In  this Letter,
we consider more physically interesting and experimentally 
accessible situation, when the nonequilibrium
distribution is unbiased and depends on the modulus of the particle
velocity, $\rho_{ne}(|v|)$, so that $\left\langle v^{2n+1} (t=0)\right\rangle \equiv0$
for any integer $n$. To be more specific, we assume for simplicity that $\rho_{ne}(|v|)$
can be represented by the Boltzmann distribution, but at a nonequilibrium
temperature $T_{ne}\neq T_{eq}$:
\begin{eqnarray}
\rho_{ne}(x,v)=\rho_{B}(x)\rho_{ne}(v),
\nonumber
\\
\rho_{ne}(v)=\sqrt{\frac{1}{2\pi}\frac{T_{eq}}{T_{ne}}}\exp\left\{ -\frac{v^{2}}{2}\frac{T_{eq}}{T_{ne}}\right\}.\label{ne}\end{eqnarray}

Such nonequilibrium distributions
over molecular velocities and angular momenta are produced, for example, in the
course of photodissociation $A+\hbar\nu\rightarrow B+\mathrm{products}$, \cite{str06,str06a,kos06d,gelkos07,dah78} which
is normally considered  as instantaneous on the timescale
of molecular translation and rotation.\cite{zew01} 
Indeed, let $\mathbf{x}_{i}$, $\mathbf{p}_{i}$, and  $m_{i}$ be the positions, the momenta, and the masses of the 
parent ($i=A$) and product ($i=B$) molecules. Since the parent molecule falls apart instantaneously,
the positions do not change, $\mathbf{x}_{B}=\mathbf{x}_{A}$. 
 As to the momenta, we can write \cite{dah78}     
$\mathbf{p}_{B}=(m_{B}/m_{A})\mathbf{p}_{A}+\mathbf{\Delta}$.
The first term here describes mapping of the parent molecule translation 
into that of the product. The quantity $\mathbf{\Delta}$
depends on the carrier frequency $\nu$ of the dissociating laser pulse and originates from the impulsive force arising due to the rupture of chemical bond(s) of the parent molecule. 
If the parent molecules possess initial equilibrium distribution over their positions and momenta, then dissociation
preserves the original distribution over the positions, $\rho_{B}(\mathbf{x})$, but produces
a nonequilibrium distribution over the momenta of photoproducts, $\rho_{ne}(\mathbf{p})$. If $\mathbf{\Delta}\ne0$,
then $\rho_{ne}(\mathbf{p})$ is a shifted Gaussian distribution.  If the laser is tuned as to make $\mathbf{\Delta}=0$, then  $\rho_{ne}(\mathbf{p})$ is the Boltzmann distribution at the nonequilibrium temperature $T_{ne}=(m_{B}/m_{A})T_{eq}$. 
In general, any reaction of the kind $A\rightarrow A'$ (photoassociation,
photoisomerization) will create a nonequilibrium distribution over
velocities of the species $A$ and $A'$, provided the processes is
rapid enough on the timescale of molecular dynamics. 

We begin with  the standard Newton equations of motion for non-interacting particles in potential $U(x)$:
 \begin{equation}
v=\dot{x},\,\,\,\dot{v}=-dU(x)/dx.\label{Ne}
\end{equation}
In the equilibrium case, initial values of positions and velocities
of the gas particles should be sampled from the Boltzmann distribution
(\ref{Boltz}). 
Since our aim is to evaluate the average velocity, 
\begin{equation}
\left\langle v(t)\right\rangle=\int dv dx \rho_{ne} (v,x) v(t),
\end{equation}
 we need to solve Eq. (\ref{Ne}) with the initial
nonequilibrium distribution (\ref{ne}).

This average velocity  $\left\langle v(t)\right\rangle $ possesses several
rather unusual properties. It is well known that any equilibrium correlation
function calculated for any Hamiltonian system is an even function
of time. This is a direct consequence of the time-inversion symmetry.
In contrast, as is easy to demonstrate,
\begin{equation}
\left\langle v(t)\right\rangle =\sum_{n=0}^{\infty}G_{n}\frac{t^{2n+1}}{(2n+1)!}\label{ser}
\end{equation}
is an odd function of time, so that $\left\langle v(t)\right\rangle =-\left\langle v(-t)\right\rangle $.
This type of behavior is caused by the special initial nonequilibrium 
preparation
of the gas at $t=0$. We can calculate the first few terms
in series (\ref{ser}) analytically: 
\begin{equation}
G_{0}=0,\,\,\, G_{1}=\frac{T_{ne}-T_{eq}}{T_{eq}}\left\langle \frac{d^{3}U(x)}{dx^{3}}\right\rangle .\label{Tay}
\end{equation}
As one may
expect, $G_{1}$ is proportional to the difference between
the equilibrium and nonequilibrium temperature, since all $G_{n}$
must vanish at equilibrium. Eq. (\ref{Tay}) shows that the external
potential must not have a symmetry axis, in order to get $\left\langle v(t)\right\rangle $$\neq0$.

To be more specific, we consider a gas trapped in the nonlinear asymmetric potential
\begin{equation}
U(x)=\omega^{2}x^{2}/2+a(x-s)^{3}+b(x-s)^{4}.
\label{u(x)}
\end{equation}
Here $\omega$ is the harmonic oscillator frequency and the parameters
$a$, $b$, and $s$ describe the nonlinear part of the potential.
One cannot analytically calculate $\left\langle v(t)\right\rangle $
for this potential. Therefore, we will compute it numerically but, first, it is instructive to
 show some results of the perturbation
theory.
Assuming that the anharmonic  part of the potential
is small and retaining linear in $a$ and $b$ contributions only,
we get 
\begin{equation}
\left\langle v(t)\right\rangle =-\frac{T_{ne}-T_{eq}}{T_{eq}}(3a-12sb)\Phi(t),\label{v(t)Fr}
\end{equation}
where
\begin{equation}
\Phi(t)=\frac{2}{3\omega^{3}}\sin(\omega t)\{1-\cos(\omega t)\}.
\label{Fi(t)Fr}
\end{equation}
So, $\left\langle v(t)\right\rangle $ exhibits a periodic
motion. Its magnitude and direction are determined by the value and sign of
$T_{ne}-T_{eq}$. Since the difference in temperatures is readily
 linked to the difference of masses in distribution
$\rho_{ne}(v)$, we see that ``loading'' and ``unloading''
of the particles (that is the increase and decrease of their masses) changes
directions of their motion, thereby creating a possibility of shuttle-like
motion.\cite{mar98} Furthermore, if we consider dissociation $A\rightarrow B+C$,
then (different) particles $B$ and $C$ are produced at different
nonequilibrium temperatures and thus induce different fluxes, allowing
for the mass separation.\cite{han99} The terms stemming from the
cubic and quartic branches of $U(x)$ enter Eq. (\ref{v(t)Fr}) with
different signs and thus lead to the opposite senses of motion. Interestingly,
the contributions cancel each other for $a=4sb$, but this is valid only
within the perturbation theory. Higher order terms ensure nonzero
$\left\langle v(t)\right\rangle $.

 \begin{figure}
\includegraphics[keepaspectratio,totalheight=7cm]{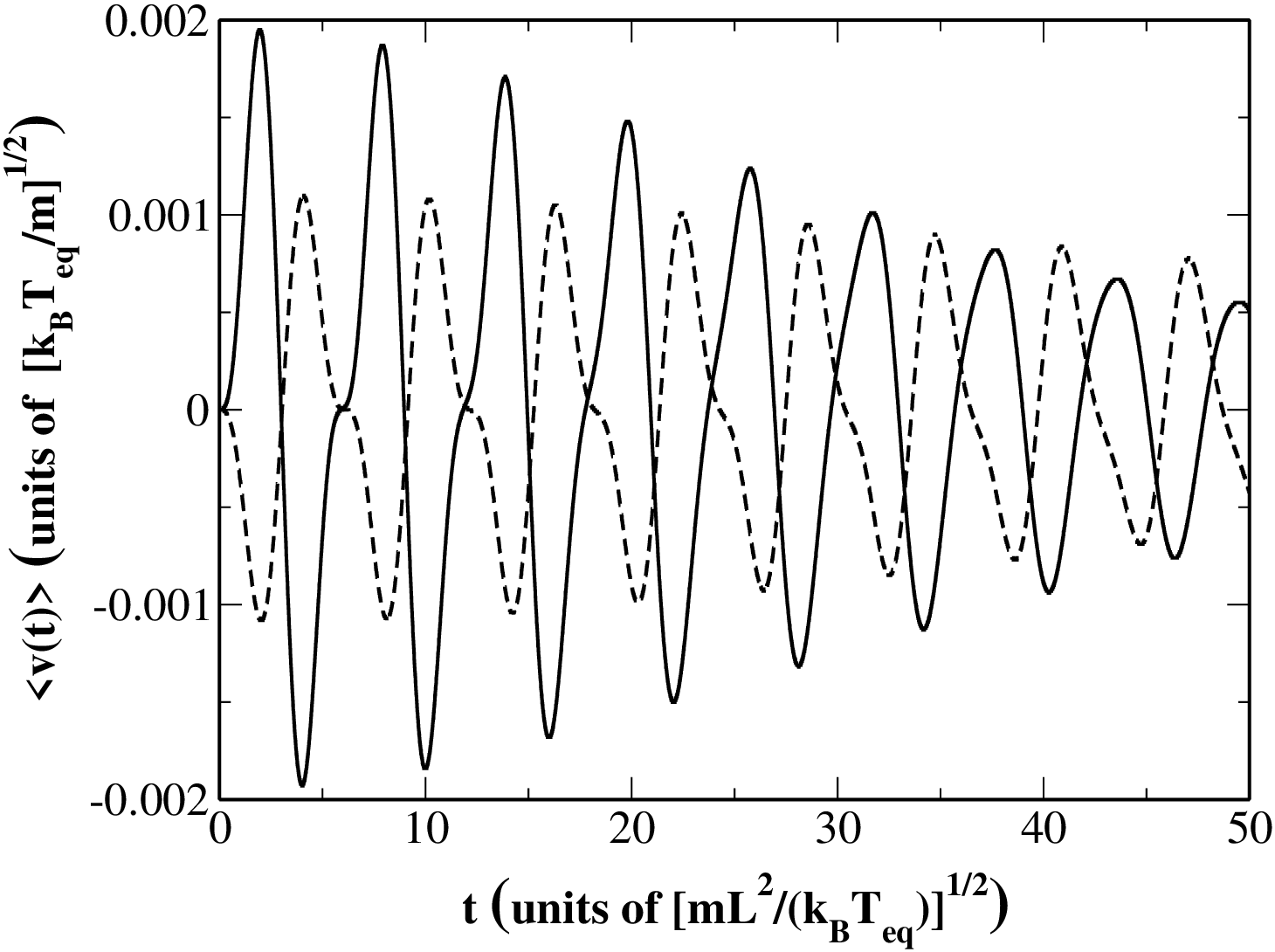}
\caption{Mean velocity $\left\langle v(t)\right\rangle $ for
ideal gas  of particles moving in the potential (\ref{u(x)}) with $\omega=1$,
$s=0.2$, $a=0.003$, and $b=0.005$. Full line corresponds to $T_{ne}=2T_{eq}$
and dashed line corresponds to $T_{ne}=T_{eq}/2$.}
\end{figure}

\begin{figure}
\includegraphics[keepaspectratio,totalheight=7cm]{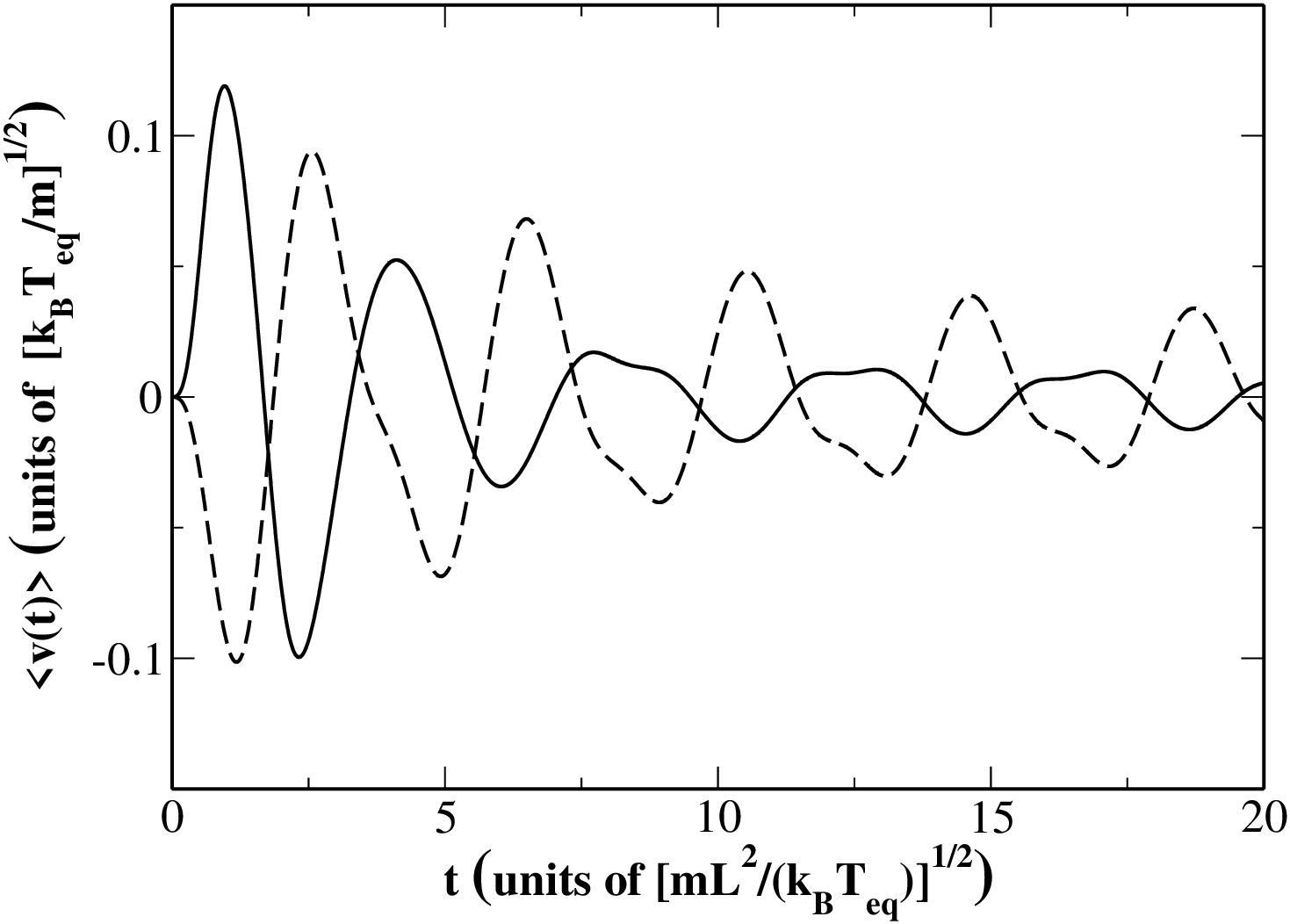}
\caption{Same as in Fig. 2 but for  $s=1$, $a=0.3$, and $b=0.5$.}
\end{figure}

We also compute $\left\langle v(t)\right\rangle $ exactly, solving
numerically equations of motions (\ref{Ne}) by Verlet algorithm for
different initial conditions sampled according to the nonequilibrium
distribution (\ref{ne}). Typical behaviors of $\left\langle v(t)\right\rangle $
are presented in Figs. 1-3. If nonlinear corrections to $U(x)$ are
small, then several first oscillations of $\left\langle v(t)\right\rangle $
in Fig. 1 are excellently described by the perturbative formulas
(\ref{v(t)Fr}) and (\ref{Fi(t)Fr}). However, the oscillations decay
in time, despite absence of dissipation. This can be understood
by switching to the action ($I$) - angle ($\varphi$) variables.
Then we can write \begin{equation}
\left\langle v(t)\right\rangle =\sum_{n=1}^{\infty}\int dId\varphi \rho_{ne}(I,\varphi)v_{n}(I)\exp\{i n(\varphi-\omega(I)t)\},\label{Ifi}
\end{equation}
$v_{n}(I)\equiv(2\pi)^{-1}\int d\varphi v(I,\varphi)\exp\{ in\varphi\}$ 
(evidently, $v_{0}(I)=0$). As distinct from the harmonic oscillator,
the frequency $\omega(I)$ of a nonlinear oscillator depends upon
the action variable and $\left\langle v(t)\right\rangle $
decays due to dephasing. 

If the the nonlinear contributions to $U(x)$ become more pronounced, then the magnitude
of $\left\langle v(t)\right\rangle $ increases (Figs. 2 and 3), but the
oscillations decay more rapidly. It is remarkable that $\left\langle v(t)\right\rangle $
can exceed the mean equilibrium thermal velocity $\sqrt{k_{B}T_{eq}/m}$ (Fig.
3). The corresponding value of the nonequilibrium temperature $T_{ne}=20T_{eq}$
is not unreasonable, e.g., for hot $CN$ fragments produced
through photodissociation of $ICN$ (see Ref. \cite{str06}). 

\begin{figure}
\includegraphics[keepaspectratio,totalheight=7cm]{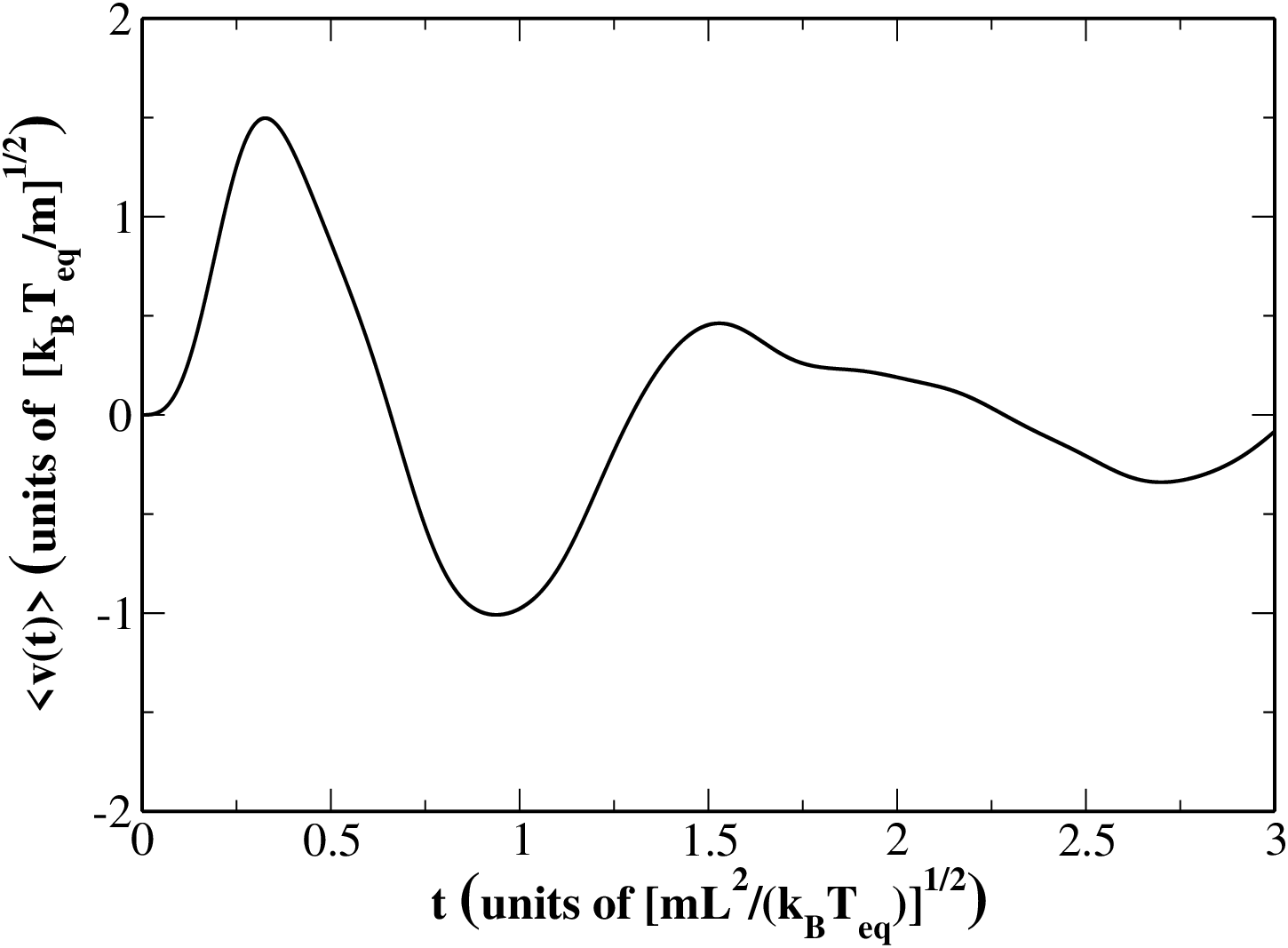}
\caption{Mean velocity $\left\langle v(t)\right\rangle $   for
ideal gas  of particles moving in the potential (\ref{u(x)}) with $\omega=1$,
$s=5$, $a=0$, and $b=5$; $T_{ne}=20T_{eq}$.}
\end{figure}

\begin{figure}
\includegraphics[keepaspectratio,totalheight=7cm]{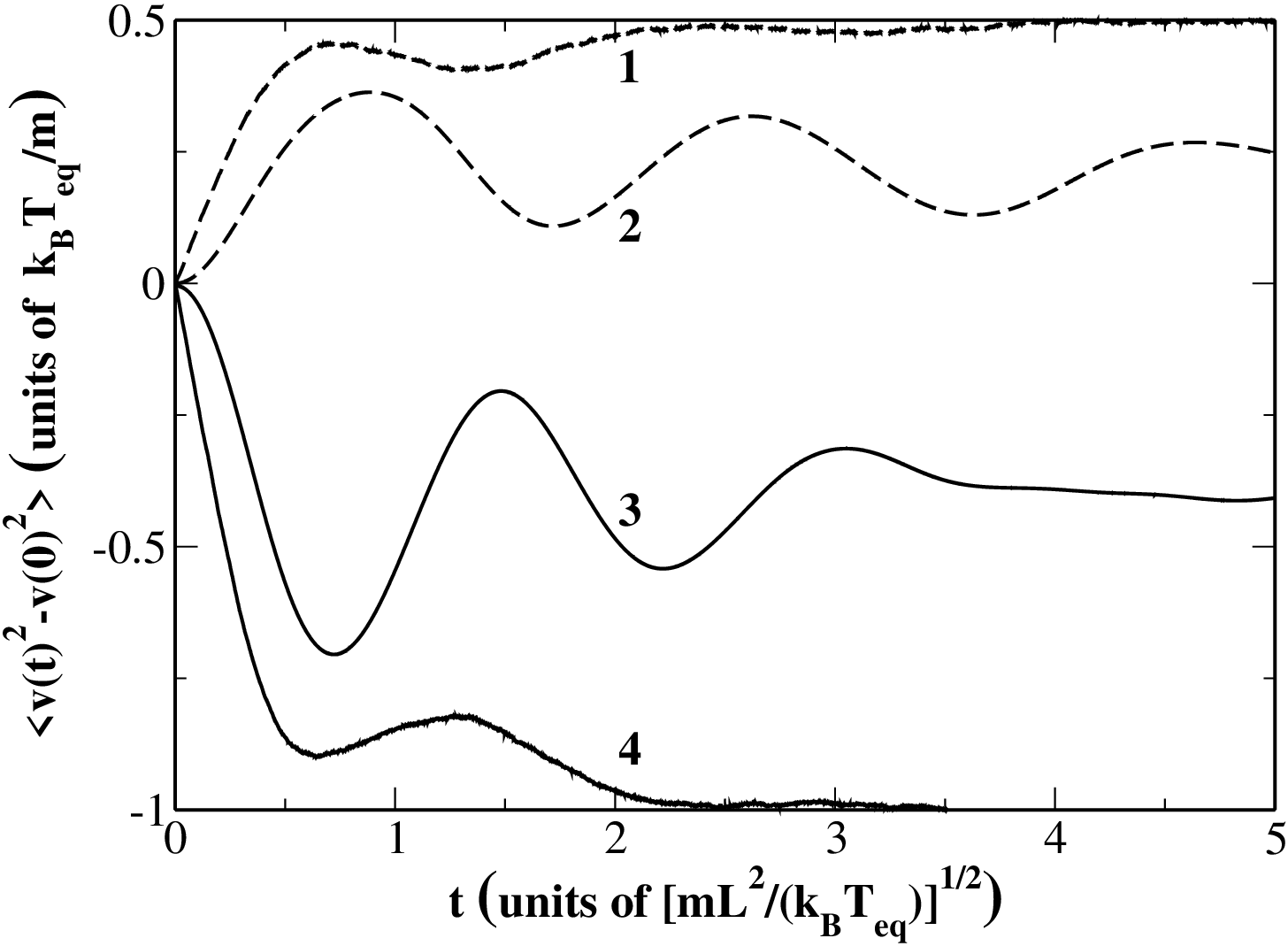}
\caption{
Total ensemble-averaged work $\left\langle v(t)^{2}-v(0)^{2}\right\rangle $
for gas of particles moving in the potential (\ref{u(x)})
with $\omega=1$, $s=1$, $a=0.3$, and $b=0.5$; for $T_{ne}=2T_{eq}$
(full line) and $T_{ne}=T_{eq}/2$ (dashed line). Curves 2 and 3
correspond to dissipation-free dynamics ($\xi=0$), curves 1 and 4
correspond to $\xi=1$.
}
\end{figure}

The quantity $\left\langle \int_{x_{1}}^{x_{2}} F(x)dx\right\rangle =\left\langle v(t_{2})^{2}-v(t_{1})^{2}\right\rangle /2$
can be regarded as the total ensemble-averaged work performed by the
gas. Evidently, this quantity is identically zero at
equilibrium. Under nonequilibrium conditions, this is nonzero oscillatory
function  (Fig. 4). The presence of oscillations means that, by choosing the appropriate  
time moments  $t_{1}$ and $t_{2}$, we can make the averaged work $\left\langle v(t_{2})^{2}-v(t_{1})^{2}\right\rangle /2$ positive,
irrespective of whether the nonequilibrium temperature $T_{ne}$ is smaller or higher than the bath temperature $T_{eq}$.  
So, our system can be regarded
as a true heat engine, which produces a cycled motion out of unbiased
nonequilibrium distributions and can be harnessed to make a useful
work. 

\begin{figure}
\includegraphics[keepaspectratio,totalheight=7cm]{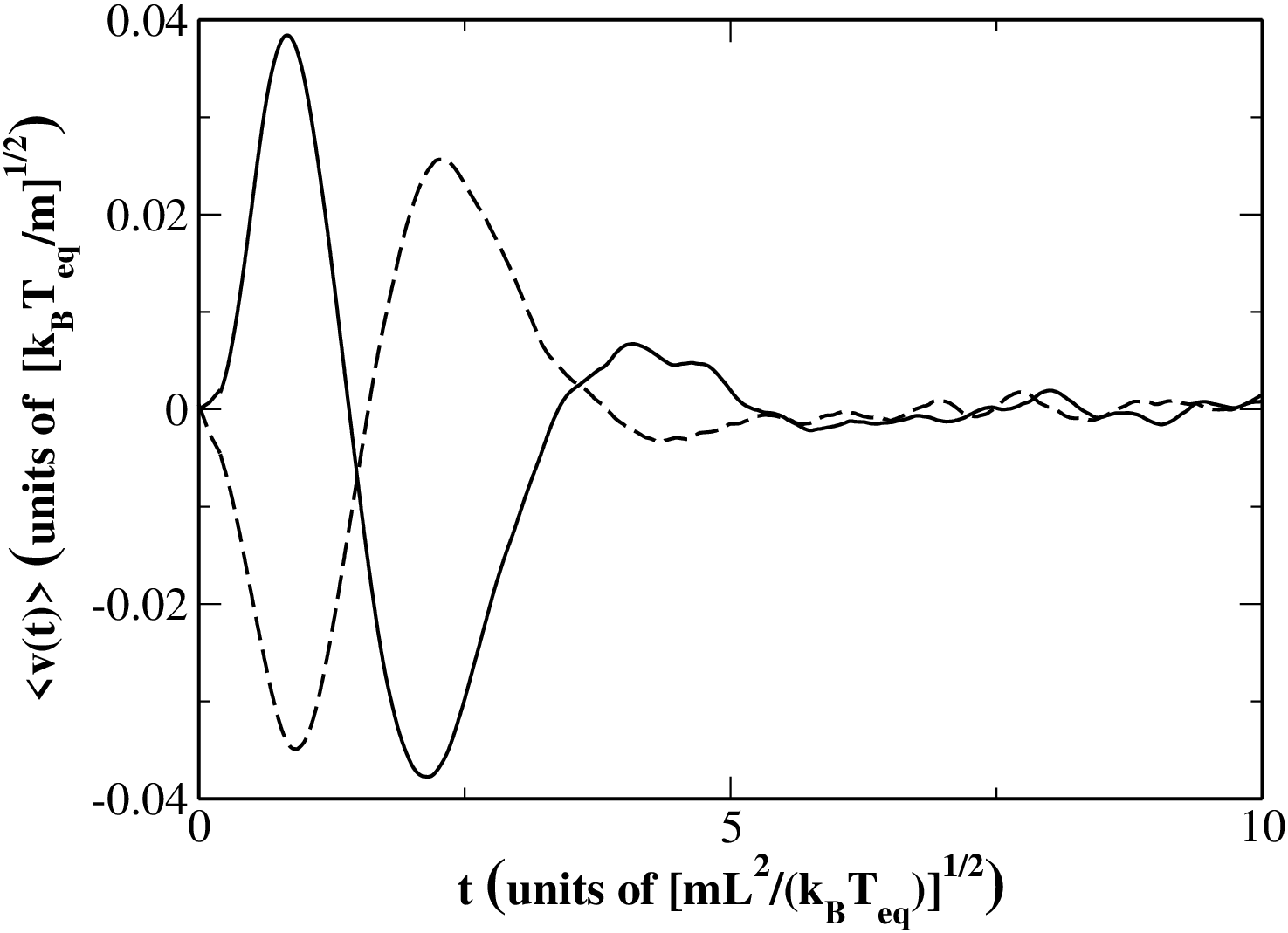}
\caption{Mean velocity $\left\langle v(t)\right\rangle $ for a moderately
damped ($\xi=1$) Brownian particle moving in the potential (\ref{u(x)})
with $\omega=1$, $s=1$, $a=0.3$, and $b=0.5$. Full line corresponds
to $T_{ne}=2T_{eq}$ and dashed line corresponds to $T_{ne}=T_{eq}/2$. }
\end{figure}

Now we turn our attention to the role of the system-bath interaction.
We thus convert the Newton equation of motion (\ref{Ne}) into 
the Langevin equation:
\begin{equation}
v=\dot{x},\,\,\,\dot{v}=-dU(x)/dx-\xi v+f(t).\label{Le}
\end{equation}
Here, $\xi$ is the friction and $f(t)$ is
 the stochastic white-noise force, which obeys the fluctuation-dissipation
relation 
$
\left\langle f(t)f(t')\right\rangle =2\xi\delta(t-t')$.
The coupling to the heat bath restores
equilibrium distribution (\ref{Boltz}), so that eventually $\left\langle v(t)\right\rangle \rightarrow0$
when $t\rightarrow\infty$. As in the bath-free case we can obtain
the perturbative solution of Eq. (\ref{Le}), which is valid up to
the terms linear in $a$ and $b$. The result is again given by Eq.
(\ref{v(t)Fr}), in which one has to put 
\begin{eqnarray}
\Phi(t)=\frac{2}{\omega^{2}\sqrt{D}}\exp\{-\frac{\xi t}{2}\}\left[\frac{\xi+\sqrt{D}}{\xi-3\sqrt{D}}\exp\{-\frac{\sqrt{D}}{2}t\}- \right.
\nonumber
\\
\left.
\frac{\xi-\sqrt{D}}{\xi+3\sqrt{D}} \exp\{\frac{\sqrt{D}}{2}t\}\right] 
-\frac{2}{D}\exp\{-\xi t\} \left[ \frac{2}{\xi-3\sqrt{D}} \times
\right.
\nonumber
\\
\left.
\exp\{\sqrt{D}t\}+\frac{2}{\xi+3\sqrt{D}}\exp\{-\sqrt{D}t\}-\frac{\xi}{\omega^{2}} \right], \,\,\,\,\,\,\,\,\,\,\,\,
\label{Fidiss}\end{eqnarray}
$D=\xi^{2}-4\omega^{2}$. In the underdamped case ($D<0$), $\Phi(t)$
consists of two contributions, which oscillate at frequencies $\sqrt{-D}$
and $\sqrt{-D}/2$, and decay $\sim\exp\{-\xi t\}$ and $\sim\exp\{-\xi t/2\}$,
correspondingly. Furthermore, there exists a monotonically decaying
contribution $\sim\xi\exp\{-\xi t\}$, which is absent in the dissipation-free
case. If $2\omega\gg\xi$, then Eq. (\ref{Fidiss}) reduces to
the damped analogue of Eq. (\ref{Fi(t)Fr})
\begin{equation}
\Phi(t)=\frac{2}{3\omega^{3}}\sin(\omega t)\exp\{-\frac{\xi}{2}t\}\left(1-\exp\{-\frac{\xi}{2}t\}\cos(\omega t)\right).
\label{under}
\end{equation}

We solve Langevin equation (\ref{Le}) numerically,
as described in.\cite{AlTi} Fig. 5 depicts $\left\langle v(t)\right\rangle $
calculated for moderate damping $\xi=1$. A comparison of Figs. 2
and 5 reveals that damping, as expected, decreases the absolute value
of $\left\langle v(t)\right\rangle $, but the effect definitely persists.
The damping does not destroy the oscillation of the ensemble-averaged work shown in Fig.4.
This means that the phenomenon of $\left\langle v(t)\right\rangle \neq0$
is quite robust and may exist in dense gases and even liquids. In
the overdamped case ($\xi\gg2\omega$, $t\gg1/\xi$) $\Phi(t)$ becomes $\sim\xi^{-3}$:
\begin{equation}
\Phi(t)=\frac{1}{\xi^{3}}\exp\{-\frac{\omega^{2}}{\xi}t\}\left(2\exp\{-\frac{\omega^{2}}{\xi}t\}-1\right).\label{over}\end{equation}
Note, however, that the question about the existence of $\left\langle v(t)\right\rangle \ne0$ is not only about the strength of friction, but also about the corresponding timescale. Since motion of any Hamiltonian system is dissipation-free for $t\ll1/\xi$, $\left\langle v(t)\right\rangle \ne0$ on that time interval.

It has not escaped our notice that, with suitably selected external periodic potential, the system represents a new type of motor which convert the nonequilibrium disparity in the velocity and coordinate distribution into directed motion. Although
both standard Brownian ratchets\cite{Rei02,han05,cis02} and our motor
use the broken symmetry of the external
potential and nonequilibrium fluctuations to produce the directed motion,  there are three important differences between them. First, the standard Brownian motors require external time-dependent (deterministic
and/or stochastic) forces lacking the detailed balance symmetry and
perpetually driving the system out of equilibrium. Alternatively, the temperature can be either driven externally or kept position-dependent. In our case, no
external driving or nonuniform temperature are necessary. The directed motion is entirely determined
by the nonequilibrium system preparation, and it is not essential
if the initial (nonequilibrium) kinetic energy $k_{B}T_{ne}/2$
is greater or smaller than the equilibrium kinetic energy, $k_{B}T_{eq}/2$.
Second, with a few notable exceptions \cite{mar98,han99,Rei02,han05,cis02,han96,han07,but98,bao00,sim02,zhe05,urb00,urb07,bro06},
the Brownian motors are normally studied in the overdamped limit.
In any case, the inertia is not a key factor ensuring the directed
motion of Brownian motors (although it is essential for the phenomenon
of absolute negative mobility \cite{han07} and sometimes is responsible
for the very occurrence of the directed transport \cite{zhe05}).
On the contrary, our motor is operating in the inertial regime,
and the effect is more pronounced in the gas phase, when friction
$\xi\ll1$ (Eq. (\ref{under})). The effect does not disappear
in the overdamped limit, but becomes very small, $\left\langle v(t)\right\rangle \sim\xi^{-3}$
(Eq. (\ref{over})). This is entirely understandable, since the higher
is the friction, the more rapidly $\rho_{ne}(v)$ relaxes to
$\rho_{B}(v)$. However, if a reversible reaction $A\longleftrightarrow A'$
takes place in our ensemble, then $\rho_{ne}(v)$ can be recovered
periodically. Third, the Brownian motors can be envisaged
as been connected to two different heat bathes.\cite{bro06,mil95,kap07}
Similarly, the "blowtorch effect" 
(i.e. a nonuniform temperature profile along the potential surface, which, in the simplest case,
can be realized within a two-bath model) can also induce a directed current.\cite{lan88,sek00}   
In our case, we can think of our motor as being connected
to kinetic and potential bathes, which are initially at different
temperatures. In this sense, it can be treated as a molecular cycled
Carnot engine. 

Note that nonlinear dissipation effects can produce 
a directed motion of the Brownian particle provided its initial velocity 
distribution is asymmetric.\cite{ply07}

The results presented in this work are not limited to a specific
Gaussian form of the nonequilibrium distribution (\ref{ne}). Any
distribution $\rho_{ne}(|v|)$ which differs from $\rho_{B}(v)$ 
gives rise to nonzero and periodic $\left\langle v(t)\right\rangle $.  
Realistic velocity distributions for dissociating molecules are 
described in papers.\cite{dah78,gelkos07,gel02}
The theory can also be straightforwardly extended to  rotational and
vibrational motion.   

Summarizing, we demonstrate that the gas of noniteracting particles trapped in
an external asymmetric potential moves with nonzero oscillatory mean
velocity $\left\langle v(t)\right\rangle $, provided its initial velocity distribution is
produced at a temperature $T_{ne}$ which differs from the equilibrium
bath temperature, $T_{eq}$. The magnitude of the effect is determined
by the value of $T_{ne}-T_{eq}$, and the direction of motion is determined
by the sign of this expression. Practically, such a nonequilibrium
distributions over molecular velocities and angular momenta are produced
in the course of photodissociation,\cite{str06,str06a,kos06d,gelkos07,dah78}
or due to any reaction of the kind $A\rightarrow A'$, provided it
is rapid enough on the timescale of particle dynamics. In molecular
dynamics simulations, we can instantaneously rescale molecular masses
(that is equivalent to the rescaling of temperature) and arrive at
the nonequilibrium distribution (\ref{ne}). The phenomenon of occurrence
of $\left\langle v(t)\right\rangle \neq0$ is robust and does not
require any careful tuning of the of the underlying parameters. 
It is not restricted to one-dimensional case and 
any asymmetric potential produces the effect, which will remain
in the quantum case, too. Furthermore, our considerations do not contain
any specific timescale and lengthscale, and are valid both for atomistic
ensembles and for collections of macroscopic objects.

\textbf{Acknowledgment.}
This work has been supported by NSF-MRSEC DMR0520471 at the University of Maryland and  by the American Chemical Society Petroleum Research Fund (44481-G6).

\end{document}